
\magnification \magstep1
\raggedbottom
\openup 4\jot
\voffset6truemm
\headline={\ifnum\pageno=1\hfill\else
\hfill{\it Gauge-averaging functionals for Euclidean Maxwell theory ...}
\hfill \fi}
\rightline {October 1993, DSF preprint 93/7 (revised version)}
\centerline {\bf GAUGE-AVERAGING FUNCTIONALS FOR}
\centerline {\bf EUCLIDEAN MAXWELL THEORY}
\centerline {\bf IN THE PRESENCE OF BOUNDARIES}
\vskip 0.3cm
\centerline {\bf Giampiero Esposito}
\vskip 0.3cm
\centerline {\it Istituto Nazionale di Fisica Nucleare}
\centerline {\it Mostra d'Oltremare Padiglione 20,
80125 Napoli, Italy;}
\centerline {\it Dipartimento di Scienze Fisiche}
\centerline {\it Mostra d'Oltremare Padiglione 19,
80125 Napoli, Italy;}
\centerline {\it Dipartimento di Fisica Teorica}
\centerline {\it Universit\`a degli Studi di Salerno}
\centerline {\it 84081 Baronissi, Italy.}
\vskip 0.3cm
\noindent
{\bf Abstract.} This paper studies
the one-loop expansion of the amplitudes
of electromagnetism about flat Euclidean backgrounds bounded
by a 3-sphere, recently considered in perturbative
quantum cosmology, by using $\zeta$-function
regularization.
For a specific choice of gauge-averaging functional,
the contributions to the full $\zeta(0)$ value owed
to physical degrees of freedom, decoupled gauge mode,
coupled gauge modes and Faddeev-Popov ghost field are
derived in detail, and alternative choices for such a
functional are also studied. This analysis
enables one to get a better understanding of different
quantization techniques for gauge fields and gravitation
in the presence of boundaries.
\vskip 0.1cm
\leftline {PACS numbers: 0370, 0460}
\vskip 100cm
\leftline {\bf 1. Introduction}
\vskip 1cm
\noindent
The way in which quantum fields respond to the presence of
boundaries is responsible for many interesting physical effects
(e.g. the Casimir effect), and plays a very important role
in quantum gravity and quantum cosmology.
In that case, the (formal) quantization
of gauge fields and gravitation via
Wick-rotated Feynman path integrals
is expressed in terms of quantum amplitudes of going from a
3-metric and a field configuration on an initial spacelike
surface to a 3-metric and a field configuration on a final
spacelike surface. Whilst mathematics enables one to understand
which compact boundary geometries do actually exist, the
methods of quantum field theory fix the boundary conditions
for scalar, fermionic and gauge fields, as well as gravitation
and corresponding ghost fields for spins $1,{3\over 2}$ and $2$.
Although the full theory via path integrals is in general
ill-defined, since there is little understanding of the measure
for quantum gravity and of the corresponding sum over all Riemannian
4-geometries, the one-loop approximations of these ill-defined
functional integrals can be evaluated in terms of well-defined
mathematical concepts [1-9]. Since one has then to study
determinants of second-order, self-adjoint, elliptic operators,
the basic tool used by theoretical physicists is the
generalized Riemann $\zeta$-function formed by the eigenvalues of
these operators as [9]
$$
\zeta(s) \equiv \sum_{n=n_{0}}^{\infty}
\sum_{m=m_{0}}^{\infty} d_{m}(n) {\lambda_{n,m}}^{-s}
\; \; \; \; .
\eqno (1.1)
$$
With our notation, $n,m$ are degeneracy labels, and
$d_{m}(n)$ is the degeneracy of the eigenvalues
$\lambda_{n,m}$, which is taken to depend only upon the
integer $n$, as happens in many physically relevant
applications (including the ones described in this paper).
The regularized $\zeta(0)$ value yields both the scaling
of the one-loop prefactor and the one-loop divergences of
physical theories.

In particular, the problem of the
one-loop finiteness of (extended) supergravity
theories in the presence of boundaries is still receiving careful
consideration in the current literature [1-9]. As emphasized
in [9-11], one can perform one-loop calculations paying
attention to: (1) S-matrix elements; (2) topological invariants;
(3) presence of boundaries. For example, in the case of pure
gravity with vanishing cosmological constant, $\Lambda=0$,
it is known that one-loop on-shell S-matrix elements are finite.
This property is also shared by $N=1$ supergravity when $\Lambda=0$,
and in that theory two-loop on-shell finiteness also holds.
However, when $\Lambda \not =0$ both pure gravity and $N=1$
supergravity are no longer one-loop finite in the sense (1) and
(2), because the non-vanishing on-shell one-loop counterterm
is given by [10]
$$
S_{(1)}={1\over {\widetilde \epsilon}}
\biggr[A \chi -{2BG\Lambda S \over 3\pi}\biggr]
\; \; \; \; .
\eqno (1.2)
$$
In (1.2) ${\widetilde \epsilon} \equiv n-4$ is
the dimensional-regularization
parameter, $\chi$ is the Euler number, $S$ is the classical
action on-shell, and one finds [9,10]:
$A={106\over 45}, B=-{87\over 10}$ for pure gravity, and
$A={41\over 24}, B=-{77\over 12}$ for $N=1$ supergravity.
Thus, $B \not =0$ is responsible for lack of S-matrix
one-loop finiteness, and $A \not =0$ does not yield topological
one-loop finiteness.

In the presence of boundaries, however, a much larger number of
counterterms can be obtained even just at one-loop order
in perturbation theory, using the extrinsic-curvature tensor
and the Ricci tensor of the boundary. Thus, if any theory of
quantum gravity can be studied from the perturbative point of
view, boundary effects play a key role in understanding whether
it has interesting and useful finiteness properties. It is
therefore necessary to analyze in detail the structure of the
one-loop boundary counterterms for fields of various spins,
and the techniques developed so far are described in detail
in [2-9]. The corresponding problems are as follows.

(i) {\it Choice of locally supersymmetric
boundary conditions} [1-4,8-9].
They involve the normal to the boundary and the field for spin ${1\over 2}$,
the normal to the boundary and the spin-${3\over 2}$ potential for
gravitinos, Dirichlet conditions for real scalar fields, magnetic or
electric field for electromagnetism, mixed boundary conditions for the
4-metric of the gravitational field (and in particular Dirichlet conditions
on the perturbed 3-metric).

(ii) {\it Quantization techniques}. One-loop amplitudes can be evaluated by
first reducing the classical theory to the physical degrees of freedom
by choice of gauge and then quantizing,
or by using the gauge-averaging method of Faddeev and Popov, or by applying
the extended-phase-space Hamiltonian path integral of Batalin, Fradkin and
Vilkovisky [2-9].

(iii) {\it Regularization techniques}.
The generalized Riemann $\zeta$-function
[12] and its regularized $\zeta(0)$ value
can be obtained by studying the
eigenvalue equations obeyed by perturbative modes, once the corresponding
degeneracies are known, or by using geometrical
formulae for one-loop counterterms which generalize well-known results for
scalar fields, but make no use of mode-by-mode eigenvalue conditions and
degeneracies [2-9,13].

Since the various quantization and regularization techniques mentioned
so far have been found to give rise to different estimates of the $\zeta(0)$
value for all spins $>0$,
it is crucial to get a better understanding of the
$\zeta(0)$ values obtained by using the
manifestly gauge-invariant quantization
techniques previously listed. The aim of this paper is to perform this
analysis in the simplest (but highly non-trivial) case, i.e. the one-loop
amplitudes of vacuum Maxwell theory
when a 3-sphere boundary is present.
One is thus led to make a
$3+1$ split of the 4-vector potential, expanding its
components on a family of 3-spheres centred on the origin as [9,14]
$$
A_{0}(x,\tau)=\sum_{n=1}^{\infty}R_{n}(\tau)Q^{(n)}(x)
\eqno (1.3a)
$$
$$
A_{k}(x,\tau)=A_{k}^{T}(x,\tau)+A_{k}^{L}(x,\tau)
\; \; \; \; \; \; \; \; {\rm for} \; {\rm all} \;
k=1,2,3
\eqno (1.3b)
$$
where $Q^{(n)}(x)$ are scalar harmonics on the 3-sphere, whereas the
transverse part $A_{k}^{T}$ and the longitudinal part $A_{k}^{L}$ are
expanded, $\forall k=1,2,3$, as
$$
A_{k}^{T}(x,\tau)=\sum_{n=2}^{\infty}f_{n}(\tau)S_{k}^{(n)}(x)
\eqno (1.4)
$$
$$
A_{k}^{L}(x,\tau)=\sum_{n=2}^{\infty}g_{n}(\tau)P_{k}^{(n)}(x)
\; \; \; \; .
\eqno (1.5)
$$
Of course, the $S_{k}^{(n)}(x)$ and $P_{k}^{(n)}(x)$ are the transverse and
longitudinal vector harmonics on $S^{3}$ respectively, and their properties
are described in detail in the appendix of [14]. Note that, strictly,
normal and tangential components of $A_{\mu}$ are only well-defined
at the 3-sphere boundary, where $\tau=a$, since a unit normal vector
field inside matching the normal to $S^3$ at the boundary is
ill-defined at the origin. As in all mode-by-mode calculations,
we are performing a {\it local} analysis, where one takes that scalar
field whose expansion on a family of 3-spheres centred on the origin
matches the $A_{0}(x,a)$ value at the boundary. Moreover, one
takes that 3-vector field whose expansion on a family of
3-spheres centred on the origin matches the $A_{k}(x,a)$ value
on $S^3$. One then has to show that an unique regular solution
of the corresponding boundary-value problem exists, although the
unit normal vector field inside is ill-defined at the origin.
It will be shown in section 3 that this is indeed the case.
[We are grateful to Dr. A. Kamenshchik for correspondence
about this problem]

Our paper is thus organized as follows.
Section 2 derives the contribution of
the physical degrees of freedom (i.e. the modes $f_{n}(\tau)$ appearing in
(1.4)) to the $\zeta(0)$ value, following [9,14]. Section 3
studies the coupled set of second-order ordinary differential equations
expressing the eigenvalue equations obeyed by the gauge modes $g_{n}$
and $R_{n}$, $\forall n \geq 2$. Section 4 derives
the contribution of the $R_{1}$-mode of (1.3a), which remains decoupled
from $R_{n}(\tau)$ and $g_{n}(\tau)$, $\forall n \geq 2$.
Section 5 studies the corresponding form of the ghost
operator, and various possible choices of the gauge-averaging term in the
Faddeev-Popov formula. Open problems and concluding remarks are
presented in section 6.
\vskip 1cm
\leftline {\bf 2. Physical degrees of freedom}
\vskip 1cm
\noindent
Within the Faddeev-Popov approach to Euclidean Maxwell theory
one deals with gauge-invariant amplitudes of the form [15]
$$
Z[g] \equiv \int {\widetilde \mu}_{1}[A]\delta[\Phi(A)] \;
{\rm det} \; \left[{\delta \Phi(A) \over \delta \epsilon}\right]
\exp \left[-\int_{M}{1\over 4}F_{\mu \nu}F^{\mu \nu}
\sqrt{{\rm det} \; g}\; d^{4}x \right]
\eqno (2.1)
$$
where $A_{\mu}$ is the 4-vector potential, $F_{\mu \nu} \equiv
\partial_{\mu}A_{\nu}-\partial_{\nu}A_{\mu}$ denotes the
electromagnetic-field tensor, and $g$ is the background 4-metric.
These amplitudes  are more conveniently re-expressed as
$$
Z[g]=\int {\widetilde \mu}_{1}[A] {\widetilde \mu}_{2}[c,c^{*}] \;
\exp{\Bigr(-{\widetilde I}_{E}\Bigr)}
\eqno (2.2)
$$
where the total Euclidean action
${\widetilde I}_{E} \equiv {\hat I}_{E}+I_{GA}+I_{gh}
\equiv I_{E}+I_{gh}$ is given by
$$
{\widetilde I}_{E}=I_{gh}+\int_{M}\left[
{1\over 4}F_{\mu \nu}F^{\mu \nu}
+{{\Bigr[\Phi(A)\Bigr]}^{2}\over 2\alpha}\right]
\; \sqrt{{\rm det} \; g} \; d^{4}x
\; \; \; \; .
\eqno (2.3)
$$
In these formulae $\Phi(A)$ is an arbitrary
gauge-averaging functional which depends on the
$U(1)$ potential $A$ and its covariant derivatives, and $\alpha$
is a positive dimensionless parameter. $I_{gh}$ is the corresponding
ghost-field action. Moreover, ${\widetilde \mu}_{1}[A]$,
${\widetilde \mu}_{2}[c,c^{*}]$,
${\rm det} \left[{\delta \Phi(A) \over
\delta \epsilon}\right]$ are a suitable measure on the
space of connections, a suitable measure for ghosts,
and the Faddeev-Popov determinant respectively. The inclusion
of gauge-averaging functionals and corresponding ghost fields
(cf section 5) is necessary to extract the volume of the gauge
group. In other words, on integrating over all field configurations
one integrates infinitely many times over the volume of the
gauge group, whereas we need to concentrate the measure over
a subset of configurations containing a single point for each
orbit of the gauge group. This is achieved using
(2.1)-(2.3).

In recent years, the case of flat Euclidean backgrounds bounded
by a 3-sphere has been studied
as the first step of a program aiming
to get a better understanding of one-loop properties of supersymmetric
field theories in the presence of boundaries [9]. As described in
section 1, for this purpose one is then led to make a $3+1$
split of the 4-vector potential as in (1.3)-(1.5). In this
section we compute the contribution of the physical degrees of
freedom $f_{n}(\tau)$ of (1.4) to the one-loop amplitudes
$Z^{(1)}$ of vacuum Maxwell theory in four-dimensions.
If the measure in the path integral is scale-invariant
(see [9,14] and references therein) such $Z^{(1)}$ amplitudes
take the asymptotic form
$$
Z^{(1)}(a) \sim W \; a^{\zeta(0)} \; e^{-I}
\eqno (2.4)
$$
where $W$ is an arbitrary constant and $a$ is the
3-sphere radius.

The physical modes $f_{n}(\tau)$ are always decoupled from the
gauge modes by virtue of the properties of the transverse vector
harmonics. The corresponding elliptic operator in the Euclidean
action is found to be [9,14]
$$
D_{n} \equiv -{1\over \tau}{d\over d\tau}\Bigr(\tau
{d\over d\tau} \Bigr)+{n^{2}\over \tau^{2}}
\; \; \; \; \; \; \; \;
{\rm for} \; {\rm all} \; n \geq 2
\eqno (2.5)
$$
whose eigenfunctions are $f_{n}(\tau)=A_{n}J_{n}
(\sqrt{E}\tau)$, $A_{n}$ being a constant [16]. Since
locally supersymmetric boundary conditions require that
either the magnetic field or the electric field should
vanish on $S^{3}$ [9,14], one has to compute the regularized
$\zeta(0)$ value for the generalized zeta-function obtained
from the eigenvalues of $D_{n}$ when $f_{n}(\tau)$ is subject
to Dirichlet conditions on $S^{3}$ (i.e. magnetic case)
or Neumann conditions on $S^{3}$ (i.e. electric case).
In the magnetic case, following the detailed analysis of
[14], one finds
$$
\zeta_{B}^{(PDF)}(0)=-{77\over 180}
\eqno (2.6)
$$
where the label (PDF) reminds us that (2.6) is the contribution
of the physical degrees of freedom to the full $\zeta(0)$. In the
electric case, following section 5.9 of [9], we begin by taking
the Laplace transform of the heat equation, where
$J_{n}(\sqrt{E}\tau)$ is replaced by the linear combination
$A_{n}I_{n}(\sigma \tau)+B_{n}K_{n}(\sigma \tau)$. The ratio
${A_{n}\over B_{n}}$ is found by requiring that
${d\over d\tau}\Bigr(A_{n}I_{n}(\sigma \tau)+B_{n}
K_{n}(\sigma \tau)\Bigr)(a)=0$,
$\forall n \geq 2$,
which takes into account the eigenvalue condition
${\dot J}_{n}(\sqrt{E}a)=0$, where $a$ is the 3-sphere
radius. Thus, the Laplace transform of the kernel of the heat
equation for spin $1$ when ${\dot f}_{n}(a)=0$,
$\forall n \geq 2$, is an infinite sum of products $G_{n}$ of
functions ${\widetilde G}_{n}$ of the type
${\widetilde G}_{n}=T\Bigr(A_{n}I_{n}(\sigma T)
+B_{n}K_{n}(\sigma T)\Bigr)$. More precisely, defining
$\tau_{<} \equiv min(\tau,\tau')$,
$\tau_{>} \equiv max(\tau,\tau')$, one finds that
$G_{n}\Bigr(\tau,\tau',\sigma^{2}\Bigr)=
{\widetilde G}_{n}\Bigr(\tau_{<},\sigma^{2}\Bigr)
{\widetilde G}_{n}\Bigr(\tau_{>},\sigma^{2}\Bigr)$, where
$$
{\widetilde G}_{n}\Bigr(\tau_{<},\sigma^{2}\Bigr)
=\tau_{<}I_{n}(\sigma \tau_{<})
\eqno (2.7a)
$$
$$
{\widetilde G}_{n}\Bigr(\tau_{>},\sigma^{2}\Bigr)
=\tau_{>}\biggr[K_{n}(\sigma \tau_{>})-
{K_{n}'(\sigma a)\over I_{n}'(\sigma a)}
I_{n}(\sigma \tau_{>})\biggr]
\; \; \; \; .
\eqno (2.7b)
$$
This implies that the free part of the heat kernel is equal to the
one found in [14], and hence does not contribute to $\zeta(0)$.
We therefore study the interacting part [9]
$$
G^{int}\Bigr(\sigma^{2}\Bigr)=-\sum_{n=2}^{\infty}
\Bigr(n^{2}-1\Bigr)
{K_{n}'(\sigma a)\over K_{n}(\sigma a)}
{I_{n}(\sigma a)\over I_{n}'(\sigma a)}
f(n;\sigma a)
\eqno (2.8)
$$
where $f(n;\sigma a)$ is the function defined in equation (4.1.18)
of [9]. Thus, we have to work out the uniform asymptotic
expansions of the various terms on the r.h.s. of (2.8)
according to the relations (4.4.13)-(4.4.22) of [9].
Setting $a=1$ for simplicity, and defining
$y \equiv {n\over \sqrt{n^{2}+\sigma^{2}}}$, this yields
$$
{K_{n}'(\sigma)\over K_{n}(\sigma)}
{I_{n}(\sigma)\over I_{n}'(\sigma)} \sim
\biggr[A_{0}(y)+{A_{1}(y)\over n}+
{A_{2}(y)\over n^{2}}+
{A_{3}(y)\over n^{3}}+...\biggr]
\eqno (2.9)
$$
where [9]
$$
A_{0}(y)=-1
\eqno (2.10)
$$
$$
A_{1}(y)=-y \Bigr(1-y^{2}\Bigr)
\eqno (2.11)
$$
$$
A_{2}(y)=-{y^{2}\over 2}{\Bigr(1-y^{2}\Bigr)}^{2}
\eqno (2.12)
$$
$$
A_{3}(y)=-{y^{3}\over 16}
\Bigr(1-y^{2}\Bigr)
\Bigr(10-68y^{2}+74y^{4}\Bigr)
\; \; \; \; .
\eqno (2.13)
$$
Thus, since $f(n;\sigma) \sim {\sqrt{n^{2}+\sigma^{2}}\over
\sigma^{2}}
\biggr[{B_{1}(y)\over n}+{B_{2}(y)\over n^{2}}
+{B_{3}(y)\over n^{3}}
+{B_{4}(y)\over n^{4}}
+... \biggr]$,
using the explicit forms of $B_{i}(y)$ appearing
in equation (4.4.12) of [9] one finds
$$
{K_{n}'(\sigma)\over K_{n}(\sigma)}
{I_{n}(\sigma)\over I_{n}'(\sigma)}f(n;\sigma)
\sim {\sqrt{n^{2}+\sigma^{2}}\over \sigma^{2}}
\biggr[{C_{1}(y)\over n}+{C_{2}(y)\over n^{2}}
+{C_{3}(y)\over n^{3}}
+{C_{4}(y)\over n^{4}}
+... \biggr]
\eqno (2.14)
$$
where [9]
$$
C_{1}(y)=-{y\over 2}\Bigr(1-y^{2}\Bigr)
\eqno (2.15)
$$
$$
C_{2}(y)={y^{2}\over 2}\Bigr(1-y^{2}\Bigr)
\Bigr(2y^{2}-1\Bigr)
\eqno (2.16)
$$
$$
C_{3}(y)=-{y^{3}\over 8}\Bigr(1-y^{2}\Bigr)
\Bigr(3-20y^{2}+21y^{4}\Bigr)
\eqno (2.17)
$$
$$
C_{4}(y)=-{y^{4}\over 16}
\Bigr(1-y^{2}\Bigr)
\Bigr(9-122y^{2}+301y^{4}-196y^{6}\Bigr)
\; \; \; \; .
\eqno (2.18)
$$
Note that additional terms in (2.14) have not been
computed since they give a contribution equal to
${\rm O}(\sqrt{t})$ ($t$ being a parameter not related to Lorentzian
time), and hence do not affect the $\zeta(0)$ value.
Thus, using (2.8) and (2.14)-(2.18) and taking the inverse
Laplace transform, one finds that the integrated heat kernel
has an asymptotic expansion as $t \rightarrow 0^{+}$ given
by [9]
$$
G^{int}(t) \sim -\sum_{n=2}^{\infty}\Bigr(n^{2}-1\Bigr)
\sum_{i=1}^{4}{\widetilde f}_{i}(n,t)
+{\rm O}(\sqrt{t})
\eqno (2.19)
$$
where [9]
$$
{\widetilde f}_{1}(n,t)=-{1\over 2}e^{-n^{2}t}
\eqno (2.20)
$$
$$
{\widetilde f}_{2}(n,t)={4\over 3}{t^{3\over 2}\over \sqrt{\pi}}
n^{2}e^{-n^{2}t}
-\sqrt{t \over \pi}e^{-n^{2}t}
\eqno (2.21)
$$
$$
{\widetilde f}_{3}(n,t)=-{3\over 8}te^{-n^{2}t}
+{5\over 4}t^{2}n^{2}e^{-n^{2}t}
-{7\over 16}t^{3}n^{4}e^{-n^{2}t}
\eqno (2.22)
$$
$$
{\widetilde f}_{4}(n,t)=-{3\over 4}
{t^{3\over 2}\over \sqrt{\pi}}e^{-n^{2}t}
+{61\over 15}{t^{5\over 2}\over \sqrt{\pi}}
n^{2}e^{-n^{2}t}
-{301\over 105}{t^{7\over 2}\over \sqrt{\pi}}
n^{4}e^{-n^{2}t}
+{392\over 945}{t^{9\over 2}\over \sqrt{\pi}}
n^{6}e^{-n^{2}t}
\; \; .
\eqno (2.23)
$$
The interacting part $G^{int}(t)$ is an even function of $n$,
and we can compute its contribution to $\zeta(0)$ using the Watson
transform defined in [9]. The poles of the integrand at $0$ and
at $\pm 1$ are excluded, since the sum over all $n$ in (2.19)
only starts from $n=2$. The poles at $\pm 1$ do not contribute, because
the integrand of the Watson transform has zeros at $\pm 1$. In our case
the constant contribution arising from the poles is given by the constant
term appearing in the inverse Laplace transform of
$-{1\over 2}{\widetilde f}(0,\sigma)$, which is equal to ${1\over 4}$.
One thus finds
$$ \eqalignno{
G^{int}(t)& \sim \biggr[
{1\over 4}+{\rm O}(\sqrt{t})-\int_{0}^{\infty}\Bigr(\rho^{2}-1\Bigr)
\sum_{i=1}^{4}{\widetilde f}_{i}(\rho,t) \; d\rho \biggr]\cr
& \sim {\sqrt{\pi}\over 8}t^{-{3\over 2}}
-{1\over 4}t^{-1}
-{55\over 256}\sqrt{\pi}t^{-{1\over 2}}
-{1\over 6}
-{1\over 90}
+{1\over 4}
+{\rm O}(\sqrt{t})
&(2.24)\cr}
$$
where the contributions $-{1\over 6}$ and $-{1\over 90}$ are owed to
(2.21) and (2.23) respectively. One thus obtains the PDF value
$$
\zeta_{E}^{(PDF)}(0)={13\over 180}
\; \; \; \; .
\eqno (2.25)
$$
\vskip 1cm
\leftline {\bf 3. Coupled gauge modes}
\vskip 1cm
\noindent
The Euclidean action is obtained multiplying
the Lorentzian action by $-i$, setting $t=-i\tau$, and bearing in
mind that $(A_{0})_{L}dt=(A_{0})_{E}d\tau$, so that the $r_{n}$-modes
appearing in the Lorentzian formulae of [14]
are related to the Euclidean
$R_{n}$-modes by $r_{n}=iR_{n}$. Moreover, for
flat Euclidean backgrounds bounded by a 3-sphere, we choose the
gauge-averaging term ${\Phi^{2}\over 2\alpha}$, where $\Phi$ is
defined as (cf (5.1))
$$
\Phi \equiv
{\partial A_{0} \over \partial \tau}
+{ }^{(3)}\nabla^{i}A_{i}
= \sum_{n=1}^{\infty}{\dot R}_{n}(\tau)Q^{(n)}(x)
-\tau^{-2}\sum_{n=2}^{\infty}g_{n}(\tau)Q^{(n)}(x) \; \; \; \; .
\eqno (3.1)
$$
Note that in (3.1) we have used the property of flat backgrounds
${ }^{(4)}\nabla^{0}A_{0}={\partial A_{0} \over \partial \tau}$,
and the relation $s^{ij}=\tau^{-2}c^{ij}$ between the contravariant
3-metric $s^{ij}$, and the contravariant 3-metric $c^{ij}$
on a unit 3-sphere. Covariant differentiation on a unit $S^{3}$,
denoted by a vertical stroke, yields
$P_{k}^{(n) \; \mid k}(x)=-Q^{(n)}(x)$ [14].
With this choice of gauge-averaging functional,
the corresponding differential operator acting on $R_{n}$-modes
will turn out to be the one-dimensional Laplace operator for
scalars if $\alpha=1$,
as we would expect in the light of the expansion (1.3a).
Note also that (3.1) does not represent the Lorentz gauge-averaging
functional (various alternative possibilities are studied in section 5).
Thus, the part $I_{E}(g,R)$ of the Euclidean action quadratic in gauge
modes is in our case [9]
$$ \eqalignno{
I_{E}(g,R)-{1\over 2\alpha}\int_{0}^{1}\tau^{3}({\dot R}_{1})^{2} \; d\tau &=
\sum_{n=2}^{\infty}\int_{0}^{1}
\left[{\tau \over 2(n^{2}-1)}{\Bigr({\dot g}_{n}-(n^{2}-1)R_{n}\Bigr)}^{2}
\right. \cr
& \left.
+{\tau \over 2\alpha}{\Bigr(-{g_{n}\over \tau}+\tau {\dot R}_{n}\Bigr)}^{2}
\right] \; d\tau
&(3.2)\cr}
$$
where we have inserted the flat-background hypothesis
$N=1$, $a(\tau)=\tau$. The physical degrees of freedom
and the ghost field decouple from (3.2).
Because we are here dealing with all degrees of freedom, we need further
boundary conditions on the modes for $A_{0}$ and the whole of $A_{k}$. For
example, we may set to zero on $S^3$ the whole of $A_{k}$: $f_{n}(1)=g_{n}(1)
=0,\; \forall n \geq 2$. This, of course,
implies the vanishing on $S^3$ of the
magnetic field {\bf B}, whereas the converse does not hold, because {\bf B}
only depends on the $f_{n}$-modes. As explained in [17], in this
case the gauge-averaging term has to vanish as well on $S^3$
by virtue of Becchi-Rouet-Stora-Tyutin (BRST) invariance. In light
of (3.1), this implies that ${\dot R}_{n}(1)=0,\; \forall n \geq 1$. The
ghost operator is then self-adjoint only if Dirichlet boundary conditions
are imposed. Viceversa, if Neumann boundary conditions are chosen for the
ghost field, remaining boundary conditions compatible with BRST invariance
are ${\dot f}_{n}(1)=0$ and ${\dot g}_{n}(1)=0,\; \forall n \geq 2$;
$R_{n}(1)=0,\; \forall n\geq 1$. This case is then called electric.

We now integrate by parts in (3.2) and use the generalized magnetic or
electric boundary conditions described above. Thus, defining
$\forall n \geq 2$ the
second-order differential operators
$$
{\hat A}_{n}(\tau) \equiv -{d^{2}\over d\tau^{2}}-{1\over \tau}
{d\over d\tau}
+{(n^{2}-1)\over \alpha \tau^{2}}
\eqno (3.3)
$$
$$
{\hat B}_{n}(\tau) \equiv {1\over \alpha}\left(-{d^{2}\over d\tau^{2}}
-{3\over \tau}{d\over d\tau}\right)+{(n^{2}-1)\over \tau^{2}}
\eqno (3.4)
$$
we find $\forall n \geq 2$ the fundamental result [9]
$$ \eqalignno{
I_{E}^{(n)}(g,R) &= {1\over 2}
\int_{0}^{1}{\tau g_{n}\over (n^{2}-1)}({\hat A}_{n}g_{n})\; d\tau
+{1\over 2}\int_{0}^{1}\tau^{3}R_{n}({\hat B}_{n}R_{n}) \; d\tau \cr
&+\left(1-{1\over \alpha}\right)\int_{0}^{1}\tau g_{n}{\dot R}_{n} \; d\tau
+\int_{0}^{1}g_{n}R_{n} \; d\tau \; \; \; \; .
&(3.5)\cr}
$$
By virtue of gauge-invariance,
we can perform the $\zeta(0)$ calculation
setting $\alpha=1$, so that the contribution of
$\tau g_{n} {\dot R}_{n}$ vanishes.
However, our problem remains a coupled one, also
with this choice.
The eigenvalues for $g_{n}$-modes and $R_{n}$-modes
can be obtained in principle from the boundary conditions and the
variational principle
$\delta \Bigr(I_{E}-\lambda J_{E}\Bigr)=0$,
where $J_{E} \equiv
{1\over 2}\int A_{\mu}A^{\mu}\sqrt{{\rm det} \; g}\;d^{4}x$.
We here make the analytic continuation to the Euclidean time variable
$\tau=it$ in computing $I_{E}$ (cf (3.2))
and $J_{E}$, and we use (1.3)-(1.5)
and the well-known properties of longitudinal vector harmonics
and scalar harmonics [14]. This leads to the
coupled system of two second-order ordinary differential equations for
arbitrary $\alpha$ and $\forall n \geq 2$ (the case $n=1$ only involves the
$R_{1}$-mode, and should be treated separately, as in section 4)
$$
{\tau \over (n^{2}-1)}\left[-\ddot g_{n}-{\dot g_{n}\over \tau}
+{(n^{2}-1)\over \alpha \tau^{2}}g_{n}\right]
+\left(1-{1\over \alpha}\right)\tau {\dot R}_{n}+R_{n}
={\lambda_{n}\over (n^{2}-1)}\tau g_{n}
\eqno (3.6)
$$
$$
\tau^{3}\left[{1\over \alpha}\left(-\ddot R_{n}-{3\over \tau}\dot R_{n}
\right)+{(n^{2}-1)\over \tau^{2}}R_{n}\right]-\tau \dot g_{n}
\left(1-{1\over \alpha}\right)+{g_{n}\over \alpha}=
\lambda_{n}\tau^{3}R_{n} \; \; .
\eqno (3.7)
$$
Now we still choose $\alpha=1$, because it enables one
to decouple much more easily the system (3.6)-(3.7).
The boundary conditions are regularity at the origin
$$
g_{n}(0)=R_{n}(0)=0 \; \; \; \; \; \; \; \;
{\rm for} \; {\rm all} \;  n \geq 2
\eqno (3.8)
$$
and magnetic conditions on $S^3$
$$
g_{n}(1)={\dot R}_{n}(1)=0 \; \; \; \; \; \; \; \;
{\rm for} \; {\rm all} \; n \geq 2
\eqno (3.9)
$$
or electric conditions on $S^3$
$$
{\dot g}_{n}(1)=R_{n}(1)=0 \; \; \; \; \; \; \; \;
{\rm for} \; {\rm all} \; n \geq 2
\; \; \; \; .
\eqno (3.10)
$$
In the $\alpha=1$ gauge we can express $R_{n}$ from (3.6) as
$$
R_{n}={\lambda_{n}\over (n^{2}-1)}\tau g_{n}+
{\tau \over (n^{2}-1)}
\left[\ddot g_{n}+{\dot g_{n}\over \tau}-{(n^{2}-1)\over \tau^{2}}g_{n}
\right]
\eqno (3.11)
$$
and its insertion into the corresponding form of (3.7) yields the
fourth-order equation
$$ \eqalignno{
0&=\left[\left(2-{3\over (n^{2}-1)}\right)\lambda_{n}\tau^{2}
-{\lambda_{n}^{2}\over (n^{2}-1)}\tau^{4}
-\Bigr(n^{2}-1\Bigr)\right]g_{n}\cr
&+2\tau \left[1-{3\lambda_{n}\tau^{2}\over (n^{2}-1)}\right]
{\dot g}_{n}
+{2\tau^{2}\over (n^{2}-1)}\biggr[
n^{2}-4 -\lambda_{n}\tau^{2}\biggr]{\ddot g}_{n}\cr
&-{6\tau^{3}\over (n^{2}-1)}g_{n}^{\rm III}
-{\tau^{4}\over (n^{2}-1)}g_{n}^{\rm IV}  \; \; \; \; .
&(3.12)\cr}
$$
Moreover, studying first the magnetic case, the relations (3.9) and
(3.11) lead to
$$
\lambda_{n}=\Bigr(n^{2}-1\Bigr)-2{\ddot g_{n}(1)\over \dot g_{n}(1)}
-{g_{n}^{\rm III}(1)\over \dot g_{n}(1)} \; \; \; \; \; \; \; \;
{\rm for} \; {\rm all} \; n \geq 2 \; \; \; \; .
\eqno (3.13)
$$
Of course, as shown by (3.6)-(3.7), the eigenvalues
$\lambda_{n}$ have dimension (length)$^{-2}$. However, in (3.13)
we have set $a=1$ for simplicity, following (3.9)-(3.10).
Hence the physical dimension does not appear explicitly.
For the solutions of the equations (3.11)-(3.12) subject to the boundary
conditions (3.8)-(3.9), an existence and uniqueness theorem holds. Thus,
denoting by $k$ an integer $\geq 0$,
in the light of the form of (3.12) we write its solution as [9]
$$
g_{n}(\tau)=\tau^{\mu}\sum_{k=0}^{\infty}a_{n,k}(n,k,\lambda_{n})
\tau^{k} \; \; \; \; .
\eqno (3.14)
$$
The insertion of (3.14) into (3.12)-(3.13), and the requirement
that $g_{n}(1)=0$,$\forall n \geq 2$, leads to a problem formulated in
purely algebraic terms. One then finds that
only half of the $a_{n,k}$ coefficients are non-vanishing
and obey very involved recurrence relations,
whereas the value of $\mu$ is obtained by solving a fourth-order algebraic
equation.

In fact, defining
$$
F(k,n,\mu)\equiv 2(k+\mu)^{2}-(n^{2}-1)-
{(k+\mu)^{2}{\Bigr((k+\mu)^{2} -1\Bigr)}\over (n^{2}-1)}
\eqno (3.15)
$$
we find
$$
F\Bigr(0,n,\mu\Bigr)a_{n,0}
=F\Bigr(1,n,\mu\Bigr)a_{n,1}=0
\eqno (3.16)
$$
$$
F\Bigr(m,n,\mu\Bigr)a_{n,m}
+\left[2-{{\Bigr(2(m+\mu)^{2}-4(m+\mu)+3\Bigr)}\over (n^{2}-1)}
\right] \lambda_{n} \; a_{n,m-2}=0
\eqno (3.17)
$$
where $m=2,3$, whereas, $\forall k \geq 4$, we have
$$ \eqalignno{
0&=F(k,n,\mu)a_{n,k}
+\left[2-{{\Bigr(2(k+\mu)^{2}-4(k+\mu)+3\Bigr)}\over (n^{2}-1)}
\right] \lambda_{n} \; a_{n,k-2}\cr
&-{\lambda_{n}^{2}\over (n^{2}-1)}a_{n,k-4} \; \; \; \; .
&(3.18)\cr}
$$
In (3.16)-(3.18), the value of $\mu$ can be obtained
from the equation $F(0,n,\mu)=0$, and bearing in mind (3.8),
which implies that only a $\mu >1$ is an acceptable value,
in the light of (3.11). In other words,
we study the fourth-order algebraic equation [9]
$$
\mu^{4}-(2n^{2}-1)\mu^{2}+(n^{2}-1)^{2}=0
\; \; \; \; .
\eqno (3.19)
$$
This equation can be easily solved setting $\mu^{2}=x$ and
studying the corresponding second-order equation for $x$.
One thus finds the four roots
$$
\mu_{+}^{(1)}=
+\sqrt{n^{2}-{3\over 4}}+{1\over 2}
\eqno (3.20)
$$
$$
\mu_{+}^{(2)}=
+\sqrt{n^{2}-{3\over 4}}-{1\over 2}
\eqno (3.21)
$$
$$
\mu_{-}^{(1)}=-\mu_{+}^{(1)}
\eqno (3.22)
$$
$$
\mu_{-}^{(2)}=-\mu_{+}^{(2)} \; \; \; \; .
\eqno (3.23)
$$
Interestingly, both $\mu_{+}^{(1)}$ and $\mu_{+}^{(2)}$ are $>1$,
$\forall n \geq 2$. They yield the desired regular solution of the
system (3.6)-(3.7).
\vskip 10cm
\leftline {\bf 4. Decoupled gauge mode}
\vskip 1cm
\noindent
In the light of (3.2), if we set $\alpha=1$ and integrate by parts
using (3.8)-(3.10), we find that the decoupled gauge mode
$R_{1}(\tau)$ obeys the eigenvalue equation
$$
{\ddot R}_{1}+{3\over \tau}{\dot R}_{1}+\mu_{1}R_{1}=0
\eqno (4.1)
$$
which is solved by $R_{1}(\tau)=H_{1}\tau^{-1}J_{1}
(\sqrt{\mu_{1}} \; \tau)$, where $H_{1}$ is a constant.
Thus, in the magnetic case, we study the eigenvalue condition
(setting $a=1$ for simplicity)
$$
J_{1}(\sqrt{\mu_{1}})-\sqrt{\mu_{1}}\; {\dot J}_{1}
(\sqrt{\mu_{1}})=0
\eqno (4.2)
$$
whereas in the electric case the corresponding eigenvalue condition
is
$$
J_{1}(\sqrt{\mu_{1}})=0
\; \; \; \; .
\eqno (4.3)
$$
If (4.2) holds, we have to perform a $\zeta(0)$ calculation with
just one perturbative mode, subject to a complicated eigenvalue
condition involving a linear combination of $J_{1}$ and
${\dot J}_{1}$. Of course, a regularization is still needed because
there are infinitely many solutions
${\hat \lambda}_{n}$ of (4.2). For
this purpose, it is convenient to use the technique described and
applied in [3,9,18]. The basic idea is as follows.

Given the $\zeta$-function at large $x$
$$
\zeta(s,x^{2}) \equiv \sum_{n=1}^{\infty}
{\Bigr({\hat \lambda}_{n}+x^{2}\Bigr)}^{-s}
\eqno (4.4)
$$
one has in four-dimensions
$$
\Gamma(3)\zeta(3,x^{2})=\int_{0}^{\infty}t^{2}e^{-x^{2}t}G(t) \;
dt \sim \sum_{q=0}^{\infty}B_{q}\Gamma \Bigr(1+{q\over 2}\Bigr)
x^{-q-2}
\eqno (4.5)
$$
where we have used the asymptotic expansion of the heat kernel
$G(t)$ for $t \rightarrow 0^{+}$, written as
$$
G(t) \sim \sum_{q=0}^{\infty}B_{q}t^{{q\over 2}-2}
\; \; \; \; .
\eqno (4.6)
$$
Such an asymptotic expansion does actually exist in the case of
Laplace operators subject to Dirichlet, Neumann or Robin boundary
conditions [9,13,19,20]. On the other hand, defining (cf (4.2))
$$
F_{1}(z) \equiv J_{1}(z)-z{\dot J}_{1}(z)
\eqno (4.7)
$$
one also has the identity
$$
\Gamma(3)\zeta(3,x^{2})=-N_{1}
{\biggr(-{1\over 2x}{d\over dx}\biggr)}^{3}
\; \log \Bigr((ix)^{-1}F_{1}(ix)\Bigr)
\eqno (4.8)
$$
where $N_{1}=1$ is the degeneracy of the problem. Thus, the comparison
of (4.5) and (4.8) can yield the coefficients $B_{q}$ and in
particular $\zeta(0)=B_{4}$, provided we carefully perform an uniform
asymptotic expansion of $F_{1}(ix)$. Now, making the analytic
continuation $x \rightarrow ix$ and then defining
${\widetilde \alpha} \equiv \sqrt{1+x^{2}}$,
one obtains the following asymptotic
expansions which are uniformly valid in the order as
$\mid x \mid \rightarrow \infty$:
$$
J_{1}(ix) \sim {(ix)\over \sqrt{2\pi}}
{\widetilde \alpha}^{-{1\over 2}}
e^{\widetilde \alpha}
e^{-\log(1+{\widetilde \alpha})}
\Sigma_{1}(1,{\widetilde \alpha}(x))
\eqno (4.9)
$$
$$
{J'}_{1}(ix) \sim {1\over \sqrt{2\pi}}
{\widetilde \alpha}^{1\over 2}
e^{\widetilde \alpha}
e^{-\log(1+{\widetilde \alpha})}
\Sigma_{2}(1,{\widetilde \alpha}(x))
\eqno (4.10)
$$
where $\Sigma_{1}(1,{\widetilde \alpha}(x))
\sim \sum_{k=0}^{\infty}
u_{k}({1\over {\widetilde \alpha}})$,
$\Sigma_{2}(1,{\widetilde \alpha}(x)) \sim
\sum_{k=0}^{\infty}v_{k}
({1\over {\widetilde \alpha}})$ [3,9,18].
Using (4.7) and (4.9)-(4.10), and defining
$C \equiv -\log(\sqrt{2\pi})$, one thus finds the uniform
asymptotic expansion
$$
\log \Bigr((ix)^{-1}F_{1}(ix)\Bigr)
\sim C -\log(1+{\widetilde \alpha})
+{1\over 2}\log(\widetilde \alpha)+{\widetilde \alpha} +
\sum_{l=1}^{\infty}\sum_{r=0}^{l}b_{lr}
{\widetilde \alpha}^{-l-2r}
\eqno (4.11)
$$
where the double sum on the right-hand side of (4.11) is
obtained by expanding in inverse powers
of $\widetilde \alpha$ the
$\log \Bigr({\Sigma_{1}\over {\widetilde \alpha}}
-\Sigma_{2}\Bigr)$. In the
light of (4.8) and (4.11), we conclude that
$\Gamma(3)\zeta(3,x^{2}) \sim \Bigr[\sigma_{1}+\sigma_{2}\Bigr]$,
where
$$
\sigma_{1} \sim {\biggr({1\over 2x}{d\over dx}\biggr)}^{3}
\; \biggr[-\log(1+{\widetilde \alpha})+{1\over 2}
\log(\widetilde \alpha)+{\widetilde \alpha} \biggr]
\eqno (4.12)
$$
$$
\sigma_{2} \sim {\biggr({1\over 2x}{d\over dx}\biggr)}^{3}
\; \sum_{l=1}^{\infty}\sum_{r=0}^{l}b_{lr}
{\widetilde \alpha}^{-l-2r}
\; \; \; \; .
\eqno (4.13)
$$
It can be easily checked that the asymptotic expansion of
$\sigma_{2}$ in (4.13) does not contribute to $\zeta(0)$,
whereas one finds
$$
\sigma_{1} \sim {3\over 8} x^{-5}-x^{-6}+{x^{-6}\over 2}
+\sum_{k=7}^{\infty}\omega_{k}x^{-k}
\eqno (4.14)
$$
which implies (cf (4.5))
$$
\zeta_{R_{1}}(0)={1\over 2} \Bigr(-{1\over 2}\Bigr)
=-{1\over 4}
\; \; \; \; .
\eqno (4.15)
$$
By contrast, when the eigenvalue condition (4.3) holds for the
$R_{1}$-mode, the square bracket on the right-hand side of
(4.12) contains $-{1\over 2}\log(\widetilde \alpha)$ rather than
${1\over 2} \log(\widetilde \alpha)$ (cf (4.9)). This leads to
$$
{\widetilde \zeta}_{R_{1}}(0)={1\over 2} \Bigr(-{3\over 2}
\Bigr)=-{3\over 4}
\eqno (4.16)
$$
since, again, the asymptotic expansion of the corresponding
$\sigma_{2}$ does not contribute to $\zeta(0)$. Note also that
our value (4.16) agrees with the result found in [21] in
the case of one perturbative mode subject to Dirichlet boundary
conditions. As explained in [22], the perturbative
calculations of [21] are correct, whereas the $\zeta(0)$
values with finitely many perturbative modes of appendix A of
[23] are incorrect.
\vskip 1cm
\leftline {\bf 5. Ghost-field contribution}
\vskip 1cm
\noindent
In section 3, we have chosen the
gauge-averaging term ${\Phi^{2}\over 2\alpha}$,
where the gauge-averaging functional $\Phi(A)$ can be written in the form
(cf [24])
$$
\Phi(A)\equiv \Phi_{1}(A)\equiv { }^{(4)}\nabla^{\mu}A_{\mu}-
K_{\; i}^{i}A_{0}
={\partial A_{0}\over \partial \tau}+{ }^{(3)}\nabla^{i}A_{i}
\eqno (5.1)
$$
where $K_{\; i}^{i}={3\over \tau}$ is the trace of the
extrinsic-curvature tensor of the boundary. This choice of $\Phi(A)$ leads
to the familiar one-dimensional Laplace operator acting on the
$R_{n}$-modes, which simplifies the $\zeta(0)$ calculation for the
coupled gauge modes and for the $R_{1}$-mode,
as shown in sections 3 and 4.
However, since $\Phi_{1}(A)$ is not the Lorentz gauge-averaging functional,
the corresponding ghost action does not involve the familiar Laplace
operator. This is proved (cf [15]) by studying the gauge transformation
$$
{ }^{\epsilon}A_{\mu} \equiv A_{\mu}+{ }^{(4)}\nabla_{\mu} \epsilon
=A_{\mu} + \partial_{\mu}\epsilon
\eqno (5.2)
$$
where the scalar $\epsilon$ is expanded on a family of 3-spheres centred
on the origin as
$$
\epsilon(x,\tau)=\sum_{n=1}^{\infty}\epsilon_{n}(\tau)Q^{(n)}(x)
\; \; \; \; .
\eqno (5.3)
$$
One thus finds
$$
\delta (\Phi_{1}(A)) \equiv
\Phi_{1}(A)-\Phi_{1}({ }^{\epsilon}A)
=\sum_{n=1}^{\infty}Q^{(n)}(x)\biggr[-{d^{2}\over d\tau^{2}}
+{(n^{2}-1)\over \tau^{2}}\biggr]\epsilon_{n}(\tau)
\; \; \; \; .
\eqno (5.4)
$$
This implies that the eigenfunctions of the ghost operator are of the
kind [16]
$$
{\widetilde \epsilon}_{n}(\tau)=\sqrt{\tau}J_{\sqrt{n^{2}-{3\over 4}}}
(\sqrt{E}\tau)
\; \; \; \; .
\eqno (5.5)
$$
More precisely, since the electromagnetic field is bosonic, the
corresponding ghost field is fermionic [15]. Its contribution to
the full $\zeta(0)$ is thus obtained changing sign to the
scalar-eigenfunctions contribution of (5.5), and then
multiplying the resulting number by two, since the ghost field
is complex.
We now have to perform a $\zeta(0)$ calculation which involves
Bessel functions of non-integer order, generalizing the technique described
in section 4. Here we show that,
although eigenvalues and eigenfunctions are
different, the $\zeta(0)$ calculation originating from (5.5) is closely
related to a standard $\zeta(0)$ calculation involving Bessel functions of
integer order. For this purpose, we study the simplest case, i.e.
when the ghost field obeys homogeneous Dirichlet conditions on $S^{3}$. This
leads to the eigenvalue condition
$$
J_{\sqrt{n^{2}-{3\over 4}}}(\sqrt{E}a)=0
\; \; \; \; \; \; \; \;
{\rm for} \; {\rm all} \; n \geq 1 \; \; \; \; .
\eqno (5.6)
$$
Following [9], our section 4 and (5.6), it is now useful to
define $\forall n \geq 1$ and at large $x$
$$
\nu \equiv + \sqrt{n^{2}-{3\over 4}}
\eqno (5.7)
$$
$$
\alpha_{\nu}(x) \equiv \sqrt{\nu^{2}+x^{2}}
=\sqrt{n^{2}-{3\over 4}+x^{2}}
\eqno (5.8)
$$
$$
\alpha_{n}(x) \equiv \sqrt{n^{2}+x^{2}}
\; \; \; \; .
\eqno (5.9)
$$
Since the generalization of the technique of section 4 to
infinitely many perturbative modes for the ghost involves defining
$\alpha_{\nu}(x)$, whereas we are only able to perform exact
calculations using $\alpha_{n}(x)$, it is also useful to evaluate
the ratio
$$
{\alpha_{\nu}(x) \over \alpha_{n}(x)} \sim \rho_{n}(x)
\sim \left[1-{3\over 8}{\Bigr(n^{2}+x^{2}\Bigr)}^{-1}
-{9\over 128}{\Bigr(n^{2}+x^{2}\Bigr)}^{-2}
+{\rm O}\Bigr({\Bigr(n^{2}+x^{2}\Bigr)}^{-3}\Bigr)\right]
\; \; \; \; .
\eqno (5.10)
$$
The asymptotic expansion (5.10) is very
useful in that it is uniform in $n$, i.e.
it holds $\forall n \geq 1$, at large $x$. A careful study of
section 7.3 of [9] shows that, if the eigenvalue condition
(5.6) holds (whose eigenvalues are positive $\forall n \geq 1$),
(4.8) and (4.11) are generalized as
$$
\Gamma(3)\zeta(3,x^{2}) \sim \Bigr[\sigma_{1}+\sigma_{2}\Bigr]
\eqno (5.11)
$$
where
$$
\sigma_{1} \sim \sum_{n=0}^{\infty}n^{2} \biggr[
-\nu x^{-6}+\nu^{2}x^{-6}\alpha_{\nu}^{-1}
+{\nu^{2}\over 2}x^{-4}\alpha_{\nu}^{-3}
+{3\over 8}\nu^{2}x^{-2}\alpha_{\nu}^{-5}
-{\alpha_{\nu}^{-6}\over 2}+{3\over 8}\alpha_{\nu}^{-5}
\biggr]
\eqno (5.12)
$$
$$
\sigma_{2} \sim -\sum_{l=1}^{\infty}\sum_{r=0}^{l}a_{lr}
\Bigr(r+{l\over 2}\Bigr)
\Bigr(r+{l\over 2}+1\Bigr)
\Bigr(r+{l\over 2}+2\Bigr)
\sum_{n=0}^{\infty}n^{2}\nu^{2r}
{\alpha_{\nu}}^{-l-2r-6}
\; \; \; \; .
\eqno (5.13)
$$
In these formulae, obtained using uniform asymptotic expansions
of Bessel functions of non-integer order, $n^{2}$ is the
degeneracy resulting from the scalar harmonics appearing in the
expansion (5.3), $\nu$ is the order of the Bessel functions
defined in (5.7), and $\alpha_{\nu}$ has been defined in
(5.8). We now re-express $\nu^{2}$ as
$\Bigr(n^{2}-{3\over 4}\Bigr)$, and
$\alpha_{\nu}(x) \sim \alpha_{n}(x)\rho_{n}(x)$ as in (5.10).
Moreover, we use the contour formula [3,9,18]
$$
\sum_{n=0}^{\infty}n^{2k}\alpha_{n}^{-2k-m}
={{\Gamma \Bigr(k+{1\over 2}\Bigr)
\Gamma \Bigr({m\over 2}-{1\over 2}\Bigr)}\over
2\Gamma \Bigr(k+{m\over 2}\Bigr)}
\; x^{1-m} \; \; \; \; \; \; \; \;
{\rm for} \; {\rm all} \; k=1,2,3,... \; \; \; \; .
\eqno (5.14)
$$
We then point out that the asymptotic expansion (5.12) can be
cast in the form
$$
\sigma_{1} \sim \Bigr[-x^{-6}I_{\infty}^{(1)}
+x^{-6}I_{\infty}^{(2)}+x^{-4}I_{\infty}^{(3)}
+x^{-2}I_{\infty}^{(4)}-I_{\infty}^{(5)}
+I_{\infty}^{(6)}\Bigr]
\eqno (5.15)
$$
where (see appendix A)
$$
I_{\infty}^{(2)}-I_{\infty}^{(1)} \equiv
-\sum_{n=0}^{\infty}n^{3}
+\sum_{n=0}^{\infty}n^{2}
\biggr[{\nu^{2}\over \alpha_{\nu}}-\Bigr(\nu -n\Bigr)
\biggr]
\eqno (5.16)
$$
$$ \eqalignno{
I_{\infty}^{(3)} & \sim {1\over 2}\sum_{n=0}^{\infty}n^{4}
\alpha_{n}^{-3} +{9\over 16}\sum_{n=0}^{\infty}n^{4}
\alpha_{n}^{-5}
+{135\over 256}\sum_{n=0}^{\infty}n^{4}\alpha_{n}^{-7}\cr
&-{3\over 8}\sum_{n=0}^{\infty}
n^{2}\alpha_{n}^{-3}
-{27\over 64}\sum_{n=0}^{\infty}n^{2}\alpha_{n}^{-5}
-{405\over 1024}\sum_{n=0}^{\infty}n^{2}\alpha_{n}^{-7}\cr
& +{1\over 2}\sum_{n=0}^{\infty}n^{4}\alpha_{n}^{-3}
{\rm O}\Bigr(\alpha_{n}^{-6}\Bigr)
-{3\over 8}\sum_{n=0}^{\infty}n^{2}\alpha_{n}^{-3}
{\rm O}\Bigr(\alpha_{n}^{-6}\Bigr)
&(5.17)\cr}
$$
$$ \eqalignno{
I_{\infty}^{(4)} & \sim {3\over 8}\sum_{n=0}^{\infty}
n^{4}\alpha_{n}^{-5}
+{45\over 64}\sum_{n=0}^{\infty}
n^{4}\alpha_{n}^{-7}
+{945\over 1024}\sum_{n=0}^{\infty}n^{4}\alpha_{n}^{-9}\cr
&-{9\over 32}\sum_{n=0}^{\infty}n^{2}\alpha_{n}^{-5}
-{135\over 256}\sum_{n=0}^{\infty}n^{2}\alpha_{n}^{-7}
-{2835\over 4096}\sum_{n=0}^{\infty}n^{2}\alpha_{n}^{-9}\cr
&+{3\over 8}\sum_{n=0}^{\infty}n^{4}\alpha_{n}^{-5}
{\rm O}\Bigr(\alpha_{n}^{-6}\Bigr)
-{9\over 32}\sum_{n=0}^{\infty}n^{2}\alpha_{n}^{-5}
{\rm O}\Bigr(\alpha_{n}^{-6}\Bigr)
&(5.18)\cr}
$$
$$ \eqalignno{
I_{\infty}^{(5)} & \equiv {1\over 2}\sum_{n=0}^{\infty}
n^{2}\alpha_{\nu}^{-6} \cr
& \sim {1\over 2}\sum_{n=0}^{\infty}n^{2}\alpha_{n}^{-6}
+{9\over 8}\sum_{n=0}^{\infty}n^{2}\alpha_{n}^{-8}
+{1\over 2}\sum_{n=0}^{\infty}n^{2}\alpha_{n}^{-6}
{\rm O}\Bigr(\alpha_{n}^{-4}\Bigr)
&(5.19)\cr}
$$
$$ \eqalignno{
I_{\infty}^{(6)} & \equiv {3\over 8}\sum_{n=0}^{\infty}
n^{2}\alpha_{\nu}^{-5} \cr
& \sim {3\over 8} \sum_{n=0}^{\infty}n^{2}\alpha_{n}^{-5}
+{45\over 64}\sum_{n=0}^{\infty}n^{2}\alpha_{n}^{-7}
+{945\over 1024}\sum_{n=0}^{\infty}n^{2}\alpha_{n}^{-9}\cr
&+{3\over 8}\sum_{n=0}^{\infty}n^{2}\alpha_{n}^{-5}
{\rm O}\Bigr(\alpha_{n}^{-6}\Bigr)
\; \; \; \; .
&(5.20)\cr}
$$
It is therefore clear, using (5.14), that the $\zeta(0)$ value
resulting from $\sigma_{1}$ and $\sigma_{2}$ is given by
${1\over 90}=-2 \Bigr(-{1\over 180}\Bigr)$
(which coincides with the $\zeta(0)$
value corresponding to the Lorentz gauge-averaging functional)
plus additional terms owed to the
second sum in (5.16), the third and fifth sum
in (5.17)-(5.18), denoted by $T_{1},T_{2},T_{3},T_{4}$,
the third sum in (5.20), denoted by $T_{5}$, and finally (5.13).
Note that $I_{\infty}^{(5)}$ defined in (5.19) does not
contribute to the additional terms. The detailed calculation yields
$$
x^{-4}T_{1} \equiv {135\over 256}x^{-4}\sum_{n=0}^{\infty} n^{4}
\alpha_{n}^{-7}
={27\over 256}x^{-6}
\eqno (5.21)
$$
$$
x^{-4}T_{2} \equiv -{27\over 64}x^{-4}
\sum_{n=0}^{\infty}n^{2}\alpha_{n}^{-5}
=-{9\over 64}x^{-6}
\eqno (5.22)
$$
$$
x^{-2}T_{3} \equiv {945\over 1024}x^{-2} \sum_{n=0}^{\infty}
n^{4}\alpha_{n}^{-9}
={27\over 512}x^{-6}
\eqno (5.23)
$$
$$
x^{-2}T_{4} \equiv -{135\over 256}x^{-2} \sum_{n=0}^{\infty}
n^{2}\alpha_{n}^{-7}
=-{9\over 128}x^{-6}
\eqno (5.24)
$$
$$
T_{5} \equiv {945\over 1024} \sum_{n=0}^{\infty}n^{2}
\alpha_{n}^{-9}
={9\over 128}x^{-6}
\; \; \; \; .
\eqno (5.25)
$$
We now focus on (5.13) and (5.16), and we first study the
asymptotic expansion (5.13), since (5.16) gives rise to
severe technical difficulties (see below).
For this purpose, we remark that,
studying for all integer values $l \in [1,\infty[,$
$r \in [1,l]$ the function (see appendix A)
$$ \eqalignno{
I_{lr}(x) & \sim \sum_{n=0}^{\infty}n^{2}\nu^{2r}
\alpha_{\nu}^{-l-2r-6} \cr
&\sim \sum_{n=0}^{\infty}
n^{2}{\Bigr(n^{2}-{3\over 4}\Bigr)}^{r}
\alpha_{n}^{-l-2r-6}
\biggr[1+A_{lr}\alpha_{n}^{-2}
+B_{lr}\alpha_{n}^{-4}
+{\rm O}\Bigr(\alpha_{n}^{-6}\Bigr)\biggr]\cr
&\sim \Bigr[I_{lr}^{(1)}+I_{lr}^{(2)}+I_{lr}^{(3)}
+I_{lr}^{(4)}\Bigr](x)
&(5.26)\cr}
$$
one finds
$$
I_{lr}^{(1)}(x) \sim \sum_{n=0}^{\infty}
n^{2}{\Bigr(n^{2}-{3\over 4}\Bigr)}^{r}
\alpha_{n}^{-l-2r-6}
\eqno (5.27)
$$
$$
I_{lr}^{(2)}(x) \sim A_{lr} \sum_{n=0}^{\infty}
n^{2}{\Bigr(n^{2}-{3\over 4}\Bigr)}^{r}
\alpha_{n}^{-l-2r-8}
\eqno (5.28)
$$
$$
I_{lr}^{(3)}(x) \sim B_{lr} \sum_{n=0}^{\infty}
n^{2}{\Bigr(n^{2}-{3\over 4}\Bigr)}^{r}
\alpha_{n}^{-l-2r-10}
\eqno (5.29)
$$
$$
I_{lr}^{(4)}(x) \sim \sum_{n=0}^{\infty}
n^{2}{\Bigr(n^{2}-{3\over 4}\Bigr)}^{r}
\alpha_{n}^{-l-2r-6}
{\rm O}\Bigr(\alpha_{n}^{-6}\Bigr)
\eqno (5.30)
$$
where $A_{lr}$ and $B_{lr}$ are coefficients which only depend
on $l$ and $r$. The case $r=0$ is easier. Using (5.13)-(5.14),
$r=0$ leads to a contribution to $\zeta(0)$ related to
$$
T_{6} \equiv -a_{10}{15\over 8}{21\over 8}
{\Gamma \Bigr({3\over 2}\Bigr)\Gamma(3)\over
2 \Gamma \Bigr({9\over 2}\Bigr)}x^{-6}
=-{3\over 8}a_{10}x^{-6}
\eqno (5.31)
$$
where ${21\over 8}$ is the coefficient of $\alpha_{n}^{-2}$ in the
asymptotic expansion of $\rho_{n}^{-7}(x)$ (see appendix A).
If the integer $r$ is $\geq 1$, one has to study (5.26)-(5.30),
where
$$
{\Bigr(n^{2}-{3\over 4}\Bigr)}^{r}
=n^{2r}-{3\over 4}rn^{2r-2}+...+rn^{2}
{\Bigr(-{3\over 4}\Bigr)}^{r-1}
+{\Bigr(-{3\over 4}\Bigr)}^{r}
\; \; \; \; .
\eqno (5.32)
$$
Inserting (5.32) into (5.27)-(5.30), and using (5.14),
a lengthy calculation yields a contribution to $\zeta(0)$ related
to (see appendix A)
$$
T_{7} \equiv \biggr[{3\over 4}a_{11}-{9\over 8}a_{11}\biggr]x^{-6}
=-{3\over 8}a_{11}x^{-6}
\; \; \; \; .
\eqno (5.33)
$$
Note that the two terms on the r.h.s. of (5.33) are due to
the asymptotic expansions of $I_{lr}^{(1)}(x)$ and $I_{lr}^{(2)}(x)$
respectively, whereas (5.29)-(5.30) do not affect the
$\zeta(0)$ value, since they do not involve $x^{-6}$.
It now remains to evaluate the contribution of (5.16).
Indeed, defining
$$
J_{\infty} \equiv \sum_{n=0}^{\infty}n^{2}\nu \biggr({\nu \over
\alpha_{\nu}}-1\biggr)
\eqno (5.34)
$$
we point out that multiplying and dividing the round bracket by
$\Bigr(\nu+\alpha_{\nu}\Bigr)$, and then adding and subtracting
$\alpha_{\nu}$ in the numerator of the corresponding expression,
one finds by virtue of (5.8) the useful identity
$$
J_{\infty}=-x^{2}\sum_{n=0}^{\infty}n^{2}
\left[{1\over \alpha_{\nu}}-{1\over \Bigr(\nu+\alpha_{\nu}\Bigr)}
\right]=J_{\infty}^{(1)}+J_{\infty}^{(2)}
\; \; \; \; .
\eqno (5.35)
$$
Moreover, (5.10) and (5.14) show that
the contribution to $\zeta(0)$ owed to $J_{\infty}^{(1)}$ is
related to
$$
T_{8}\equiv -{27\over 128}x^{-4}\sum_{n=0}^{\infty}
n^{2}\alpha_{n}^{-5}=-{9\over 128}x^{-6}
\; \; \; \; .
\eqno (5.36)
$$
A further contribution is owed to
$$
T_{9} \equiv {1\over 120}x^{-6}
\eqno (5.37)
$$
originating from $\sum_{n=0}^{\infty}n^{3}$ in (5.16).
However, we do not yet know how to deal properly with
the divergent sum
$$
J_{\infty}^{(2)} \equiv x^{2}\sum_{n=0}^{\infty}
{n^{2}\over \Bigr(\nu+\alpha_{\nu}\Bigr)}
\; \; \; \; .
\eqno (5.38)
$$

We should now add up the numerical coefficients appearing in
(5.21)-(5.25), (5.31), (5.33), (5.36)-(5.37),
divide them by two, and finally
multiply by $-2$ since the ghost is fermionic and complex. This
leads to the following {\it partial} contribution to the
$\zeta(0)$ value for the ghost field:
$$
\zeta_{gh}^{(I)}(0)={1\over 90}-{9\over 512}
+{3\over 8}\Bigr(a_{10}+a_{11}\Bigr)
+{9\over 128}-{1\over 120}
={1\over 360}+{11\over 512}
\eqno (5.39)
$$
where ${1\over 90}$ is added for the reasons described following
(5.20), and we have used the values $a_{10}={1\over 8},$
$a_{11}=-{5\over 24}$ appearing in equation (26) of [18].

By contrast, if the Lorentz gauge-averaging functional is chosen, one finds
$$
\Phi(A) \equiv \Phi_{2}(A) \equiv { }^{(4)}\nabla^{\mu}A_{\mu}
={\partial A_{0}\over \partial \tau}+
{ }^{(4)}\nabla^{i}A_{i}
\eqno (5.40)
$$
which implies
$$
\delta(\Phi_{2}(A)) \equiv
\Phi_{2}(A)-\Phi_{2}({ }^{\epsilon}A)
=\sum_{n=1}^{\infty}Q^{(n)}(x)\biggr[-{d^{2}\over d\tau^{2}}
-{3\over \tau}{d\over d\tau}+{(n^{2}-1)\over \tau^{2}}\biggr]
\epsilon_{n}(\tau)
\eqno (5.41)
$$
where we have used the property ${ }^{(4)}\nabla_{i} \epsilon=
{ }^{(3)}\nabla_{i} \epsilon = \epsilon_{\mid i}=
\partial_{i}\epsilon$, $\forall i=1,2,3$. Thus, as we anticipated, the
familiar one-dimensional Laplace operator appears in the ghost action, so
that the ghost contributions to the full $\zeta(0)$ value are more easily
computed as $-2 \left(-{1\over 180}\right)$ and
$-2 \left({29\over 180}\right)$ in the Dirichlet and Neumann cases,
respectively. However, if $\Phi_{2}(A)$ is chosen as gauge-averaging
functional, the form of the action quadratic in the gauge modes becomes
$\forall n \geq 2$
$$ \eqalignno{
I_{E}^{(n)}(g,R)&={1\over 2}\int_{0}^{1}{\tau g_{n}\over (n^{2}-1)}
\biggr[-{d^{2}g_{n}\over d\tau^{2}}-{1\over \tau}{dg_{n}\over d\tau}
+{(n^{2}-1)\over \alpha \tau^{2}}g_{n}\biggr]\; d\tau \cr
&+{1\over 2}\int_{0}^{1}\tau^{3}R_{n}
\left[{1\over \alpha}\biggr(-{d^{2}R_{n}\over d\tau^{2}}+{3\over \tau}
{dR_{n}\over d\tau}\biggr)
+\Bigr(n^{2}-1+{9\over \alpha}\Bigr){R_{n}\over \tau^{2}}\right]
\; d\tau \cr
&+\Bigr(1-{1\over \alpha}\Bigr)\int_{0}^{1}\tau g_{n}{\dot R}_{n}
\; d\tau
+\Bigr(1-{3\over \alpha}\Bigr)\int_{0}^{1}g_{n}R_{n} \; d\tau \cr
&-{\Bigr[\tau g_{n}R_{n}\Bigr]}_{0}^{1}
+{1\over 2\alpha}{\Bigr[\tau^{3}{\dot R}_{n}R_{n}\Bigr]}_{0}^{1}
\; \; \; \; .
&(5.42)\cr}
$$
Thus, the second-order differential operator acting on $R_{n}$-modes is no
longer the one-dimensional Laplace operator for scalars, and the calculation
becomes more involved. For example, if we set $\alpha=1$, the contribution
of $R_{1}(\tau)$ to $\zeta(0)$ involves a Bessel function of order
$\sqrt{13}$. Moreover, a non-vanishing boundary term
$I_{B}^{(n)} \equiv {a^{3}\over 2\alpha}{\dot R}_{n}(a)R_{n}(a)=
-{3a^{2}\over 2\alpha}R_{n}^{2}(a)$
survives in the action, if the whole
functional $\Phi_{2}(A)$ is required to vanish on the boundary in the
magnetic case (cf [17]). Thus,
one has to add to the action a boundary
term equal to $-I_{B}^{(n)}$,
if the whole of $\Phi_{2}(A)$ is set to zero on $S^{3}$.

Of course, since the theory is gauge-invariant, {\it infinitely} many other
choices for $\Phi(A)$ (but not all choices)
are still possible. A very relevant class of choices
can be cast in the form
$$
\Phi^{(b)}(A) \equiv { }^{(4)}\nabla^{\mu}A_{\mu}+bK_{\; i}^{i}A_{0}
\eqno (5.43)
$$
where $b$ is a real number. With our parametrization, $b=-1$ leads to
$\Phi_{1}(A)$, and $b=0$ leads to $\Phi_{2}(A)$.
Note that, even if we set $\alpha=1$, it does not seem possible to
decouple gauge modes using ${\Phi^{2}\over 2\alpha}$ {\it and} obtain
a well-defined ghost action, since the decoupling of $g_{n}$ and $R_{n}$,
$\forall n \geq 2$, is then obtained setting
$$
\Phi(A)  \equiv \Phi_{3}(A) \equiv
\sum_{n=2}^{\infty}
\sqrt{\left[{\biggr(-{g_{n}\over \tau^{2}}+{\dot R}_{n}\biggr)}^{2}
+{2\over \tau^{2}}{d\over d\tau}(g_{n}R_{n})\right]} \; Q^{(n)}(x)
\; \; \; \; .
\eqno (5.44)
$$
However, the ghost action should be derived by functionally
differentiating the infinite sum of square roots on the right-hand
side of (5.44) as in (5.4) and (5.41), and this does not lead
to a linear, second-order differential operator.
This is why we believe that the coupling of gauge modes is an intrinsic
property of problems with boundaries, as well as the choice of
gauge-averaging functionals of the form (5.43), which all reduce to the
Lorentz choice in the absence of boundaries. Note that the work in this
section supersedes earlier work appearing in section 6.5 of [9],
where the ghost-field operator (cf (5.4)) was not derived.

In light of (2.6), (4.15) and (5.39), the full $\zeta(0)$ value
for vacuum Euclidean Maxwell theory
in the case of magnetic boundary conditions on
$S^{3}$ takes the form
$$
\zeta(0)= \zeta_{B}^{(PDF)}(0)+\zeta_{R_{1}}(0)
+\zeta_{GM}(0)+\zeta_{gh}(0)
=-{243\over 360}+{11\over 512}+\zeta_{GM}(0)
+\zeta_{gh}^{(II)}(0)
\eqno (5.45)
$$
where $\zeta_{GM}(0)$ and $\zeta_{gh}^{(II)}(0)$ are the as yet
unknown contributions to $\zeta(0)$ arising from coupled gauge
modes (section 3) and from (5.38) respectively. We have been
unable to evaluate $\zeta_{GM}(0)$
since we do not know explicitly the
uniform asymptotic expansion as $\lambda_{n} \rightarrow \infty$
of the power series in (3.14), which is not (obviously) related
to well-known special functions (see appendix B).
Moreover, the regularized
contribution $\zeta_{gh}^{(II)}(0)$ of (5.38) to $\zeta(0)$
involves $\nu \equiv + \sqrt{n^{2}-{3\over 4}}$,
which is a source of
complication. However, it should be emphasized that all
divergences are only {\it fictitious}, since the starting point
for the derivation of (5.12) is the identity [9]
$$
{\biggr({1\over 2x}{d\over dx}\biggr)}^{3}
\log \biggr({1\over {\nu + \alpha_{\nu}}}\biggr)
={\Bigr(\nu+\alpha_{\nu}\Bigr)}^{-3}
\biggr[-\alpha_{\nu}^{-3}-{9\over 8}\nu \alpha_{\nu}^{-4}
-{3\over 8}\nu^{2}\alpha_{\nu}^{-5}\biggr]
\; \; \; \; .
\eqno (5.46)
$$
This proves that by summing over all integer values of
$n$ from $0$ to $\infty$ one gets a convergent series.

In this section we have not studied the case of Neumann boundary
conditions for the ghost field, i.e. the electric case. This
complicated calculation may be, by itself, the object of
another paper. However, interestingly, in
light of (2.6), (2.25) and (4.15)-(4.16) one finds
$$
\zeta_{B}^{(PDF)}(0)+\zeta_{R_{1}}(0)
=\zeta_{E}^{(PDF)}(0)+{\widetilde \zeta}_{R_{1}}(0)
=-{61\over 90}
\; \; \; \; .
\eqno (5.47)
$$
In other words, if the gauge-averaging functional of (5.1)
is chosen, physical degrees of freedom and decoupled gauge mode
give the same partial contribution to the full $\zeta(0)$,
i.e. $-{61\over 90}$, both in the magnetic and in the electric
case.
\vskip 1cm
\leftline {\bf 6. Concluding remarks and open problems}
\vskip 1cm
\noindent
One-loop quantum cosmology may add further evidence in favour
of different approaches to quantizing gauge theories being
inequivalent [2-9,25-34]. Studying flat Euclidean backgrounds
bounded by a 3-sphere, for vacuum Maxwell theory the PDF method
yields $\zeta(0)=-{77\over 180}$ and $\zeta(0)={13\over 180}$
in the magnetic and electric cases respectively [6,9,14],
whereas the {\it indirect} method (by this we mean
that one-loop amplitudes are expressed using the
boundary-counterterms technique and evaluating the various
coefficients in a covariant way as in [2,17]) was found to yield
$\zeta(0)=-{38\over 45}$ in both cases in [2]. For $N=1$
supergravity, the PDF method yields {\it partial} cancellations between
spin $2$ and spin ${3\over 2}$ [7-9], whereas the {\it indirect}
method yields a one-loop amplitude which is even more
divergent than in the pure-gravity case [2]. Finally, for pure
gravity, the PDF method yields $\zeta(0)=-{278\over 45}$ in the
Dirichlet case, whereas the {\it indirect} method
yields $\zeta(0)=-{803\over 45}$ [2,9]. Moreover, within the PDF
method, it is possible to set to zero on $S^3$ the linearized
magnetic curvature. This yields a well-defined one-loop
calculation, and the corresponding $\zeta(0)$ value is
${112\over 45}$ [9]. By contrast, using the Faddeev-Popov
formula, magnetic boundary conditions for pure gravity are
ruled out [2]. Interestingly, recent work in [35] seems to
add evidence in favour of {\it direct} $\zeta(0)$ calculations
being correct. The authors of [35] have shown that different
formulae for $\zeta(0)$ previously obtained in [8] for
Majorana and Dirac fermions on the part of a de Sitter sphere
bounded by a 3-sphere, with local and spectral boundary
conditions, have the same limiting value in the case of a
full sphere. This value coincides with the covariant one
obtained by the method in [2,13,17]. Moreover, all these
expressions are found to give the same results in the case
of a hemisphere [35]. The authors of [35] have also suggested
that $3+1$ splits of the kind considered in many papers,
including Eqs. (1.3a)-(1.3b) of our paper, might be the
reason of the discrepancies for higher-spin $\zeta(0)$ values
found using covariant and non-covariant methods. However, there
is not yet a proof of this statement, and in the case of real
scalar fields subject to Neumann (or Dirichlet) conditions on a
3-sphere boundary, $\zeta(0)$ values obtained from various
methods coincide, although the boundary 3-geometry is the
same as in higher-spin calculations.

It is therefore necessary to get a better
understanding of the manifestly gauge-invariant formulae for
one-loop amplitudes used so far in the literature, by performing
a mode-by-mode analysis of the eigenvalue equations, rather
than relying on general formulae which contain no explicit information
about degeneracies and eigenvalue conditions.
This detailed analysis has been attempted here in the simplest
case, i.e. vacuum Maxwell
theory at one-loop about a flat Euclidean background bounded
by a 3-sphere. Our results are here summarized for the sake
of clarity:

(1) In the light of (2.6), (2.25)
and (4.15)-(4.16) the physical degrees of freedom,
and the decoupled gauge mode, give a contribution to the
full $\zeta(0)$ equal to $-{61\over 90}$ both in the magnetic and in the
electric case (this important property had not been realized in
section 6.5 of [9]),
if the gauge-averaging functional $\Phi_{1}(A)$ of (5.1) is chosen.
Since in [2] it was found that the full $\zeta(0)$ values for spin
$1$ are equal in the magnetic and electric cases, it appears
relevant that also our partial contributions to the full $\zeta(0)$
coincide in these two cases for the spin-1 problem about flat
Euclidean backgrounds.

(2) Remaining gauge modes $g_{n}$ and $R_{n}$ always obey a coupled system
of linear, second-order ordinary differential equations,
$\forall n \geq 2$. The solution of such a system
corresponding to $\Phi_{1}(A)$
has been given in section 3 and appendix B.

(3) If $\Phi_{1}(A)$ is chosen, the ghost eigenfunctions
involve Bessel functions of non-integer order. The corresponding
contribution to the full $\zeta(0)$ can be obtained using the method of
section 5. Such a technical point appears interesting, since to our
knowledge no previous mode-by-mode analysis for the ghost is
appearing in the literature in the case of Bessel functions of
non-integer order.

It now remains to evaluate the contribution
to the full $\zeta(0)$ of the divergent sum in (5.38),
and the uniform asymptotic expansion of $g_{n}$-
and $R_{n}$-modes as $\lambda_{n} \rightarrow \infty$ at the end of
section 3. Unfortunately,
the generalization of the method described in
[5-8] is highly non-trivial. By contrast, a
simpler form of the ghost eigenfunctions is obtained using the Lorentz
gauge-averaging functional $\Phi_{2}(A)$. However, this leads to a further
complication of the calculations involving gauge modes, since the
decoupled mode $R_{1}(\tau)$ involves a Bessel function of order
$\sqrt{13}$ (this implies a contribution to $\zeta(0)$
proportional to $\sqrt{13}$, which we find very puzzling),
and coupled gauge modes require the addition to the action,
in the magnetic case, of a boundary term equal to
${3a^{2}\over 2\alpha}\sum_{n=2}^{\infty}R_{n}^{2}(a)$.

Thus, although some evidence exists that different $\zeta(0)$ values
for gauge fields in the presence of boundaries are due to
inequivalent quantization techniques [9], the most important check,
i.e. the mode-by-mode analysis of eigenvalue equations for gauge
modes and ghost fields, remains a very difficult problem. We hope
that our paper, through its detailed (although incomplete) analysis,
may contribute to shed new light on this longstanding problem in
quantum field theory.
\vskip 1cm
\leftline {\bf Acknowledgments}
\vskip 1cm
\noindent
I am much indebted to
Bruce Allen, Andrei Barvinsky,
Peter D'Eath, Gary Gibbons, Chris Isham, Alexander Kamenshchik,
Jorma Louko, Ian Moss, Stephen Poletti
and Peter van Nieuwenhuizen for enlightening conversations
or correspondence.
Anonymous referees made comments which led to a substantial
improvement of the original manuscript.
I am also grateful to Professor Abdus Salam, the International
Atomic Energy Agency and UNESCO for hospitality at the International
Centre for Theoretical Physics, and to Professor Dennis Sciama
for hospitality at the SISSA of Trieste, during the early
stages of this work.
Last, but not least, the stimulating atmosphere of the July 1992
Les Houches Summer School on Gravitation and Quantizations has
played a key role in preparing this paper.
\vskip 1cm
\leftline {\bf Appendix A}
\vskip 1cm
\noindent
The derivation of (5.17)-(5.20),
(5.31), (5.33) and (5.36) has been
obtained using the following asymptotic expansions:
$$
\rho_{n}^{-1}(x) \sim
1+{3\over 8}\alpha_{n}^{-2}+{27\over 128}\alpha_{n}^{-4}
+{\rm O}\Bigr(\alpha_{n}^{-6}\Bigr)
\eqno (A.1)
$$
$$
\rho_{n}^{-2}(x) \sim 1 +{3\over 4}\alpha_{n}^{-2}
+{9\over 16}\alpha_{n}^{-4}
+{\rm O}\Bigr(\alpha_{n}^{-6}\Bigr)
\eqno (A.2)
$$
$$
\rho_{n}^{-3}(x) \sim
1+{9\over 8}\alpha_{n}^{-2}+{135\over 128}\alpha_{n}^{-4}
+{\rm O}\Bigr(\alpha_{n}^{-6}\Bigr)
\eqno (A.3)
$$
$$
\rho_{n}^{-4}(x) \sim 1+{3\over 2}\alpha_{n}^{-2}
+{27\over 16}\alpha_{n}^{-4}
+{\rm O}\Bigr(\alpha_{n}^{-6}\Bigr)
\eqno (A.4)
$$
$$
\rho_{n}^{-5}(x) \sim
1+{15\over 8}\alpha_{n}^{-2}+{315\over 128}\alpha_{n}^{-4}
+{\rm O}\Bigr(\alpha_{n}^{-6}\Bigr)
\eqno (A.5)
$$
$$
\rho_{n}^{-6}(x) \sim
1+{9\over 4}\alpha_{n}^{-2}+{27\over 8}\alpha_{n}^{-4}
+{\rm O}\Bigr(\alpha_{n}^{-6}\Bigr)
\eqno (A.6)
$$
$$
\rho_{n}^{-7}(x) \sim
1+{21\over 8}\alpha_{n}^{-2}+{567\over 128}\alpha_{n}^{-4}
+{\rm O}\Bigr(\alpha_{n}^{-6}\Bigr)
\eqno (A.7)
$$
$$
\rho_{n}^{-8}(x) \sim
1+3\alpha_{n}^{-2}
+{45\over 8}\alpha_{n}^{-4}
+{\rm O}\Bigr(\alpha_{n}^{-6}\Bigr)
\eqno (A.8)
$$
$$
\rho_{n}^{-9}(x) \sim
1+{27\over 8}\alpha_{n}^{-2}
+{891\over 128}\alpha_{n}^{-4}
+{\rm O}\Bigr(\alpha_{n}^{-6}\Bigr)
\; \; \; \; .
\eqno (A.9)
$$
Note that these expansions are valid uniformly in the integer $n$,
$\forall n \geq 1$, as $\mid x \mid \rightarrow \infty$.
They are obtained using repeatedly (5.10) and the well-known
expansion of $(1+Y)^{-1}$ as $Y \rightarrow 0$.
\vskip 1cm
\leftline {\bf Appendix B}
\vskip 1cm
\noindent
Following section 3, coupled gauge modes can be written as
$$
g_{n}^{(j)}(\tau)=\sum_{k=0}^{\infty}a_{n,k}^{(j)}
\Bigr(n,k,\lambda_{n}^{(j)}\Bigr)\tau^{k+\mu}
\eqno (B.1)
$$
$$
R_{n}^{(j)}(\tau)=\sum_{k=0}^{\infty}b_{n,k}^{(j)}
\Bigr(n,k,\lambda_{n}^{(j)}\Bigr)\tau^{k+\mu-1}
\; \; \; \; .
\eqno (B.2)
$$
The label $j$ is introduced because, for each integer
value of $n \geq 2$, there is a whole family
$\Bigr \{\lambda_{n}^{(j)} \Bigr \}$ of eigenvalues
labelled by the integer $j$, say. They are solutions
of the equation $g_{n}(a)=0$, and their degeneracy
$d_{j}(n)=n^{2}$, $\forall j \geq 1$ and
$\forall n \geq 2$ [14]. Now, defining
$\forall k \geq 2$ and $\forall n \geq 2$
$$
G(k,n,\mu) \equiv 2
-{{\Bigr(2(k+\mu)^{2}-4(k+\mu)+3\Bigr)}\over
(n^{2}-1)}
\eqno (B.3)
$$
one finds $\forall j \geq 1$
$$
{a_{n,2}^{(j)}\over a_{n,0}^{(j)}}=
-{G(2,n,\mu)\over F(2,n,\mu)}\lambda_{n}^{(j)}
\eqno (B.4)
$$
$$
F(k,n,\mu)a_{n,k}^{(j)}+G(k,n,\mu)\lambda_{n}^{(j)}
a_{n,k-2}^{(j)}-{{\Bigr(\lambda_{n}^{(j)}\Bigr)}^{2}
\over (n^{2}-1)}a_{n,k-4}^{(j)}=0
\; \; \; \; \; \; \; \;
\forall k \geq 4
\eqno (B.5)
$$
$$
b_{n,0}^{(j)}=\biggr({\mu^{2}\over (n^{2}-1)}-1\biggr)
a_{n,0}^{(j)}
\eqno (B.6)
$$
$$
b_{n,k}^{(j)}={\lambda_{n}^{(j)}\over (n^{2}-1)}
a_{n,k-2}^{(j)}+\biggr({(k+\mu)^{2}\over (n^{2}-1)}-1
\biggr)a_{n,k}^{(j)}
\; \; \; \; \; \; \; \;
\forall k \geq 2
\eqno (B.7)
$$
$$
a_{n,k}^{(j)}=b_{n,k}^{(j)}=0
\; \; \; \; \; \; \; \;
\forall k =(2m+1)
\; \; \; \; \; \; \; \;
m=0,1,2,...
\; \; \; \; .
\eqno (B.8)
$$
Moreover, setting the 3-sphere radius $a$ to $1$ for simplicity,
magnetic boundary conditions (i.e.
$g_{n}(1)={\dot R}_{n}(1)=0$) lead to
$$
\sum_{k=0}^{\infty}a_{n,k}^{(j)}=0
\eqno (B.9)
$$
$$
\lambda_{n}^{(j)}=\Bigr(n^{2}-1\Bigr)-\mu(3\mu-2)
-{{\sum_{k=0}^{\infty}k^{3}a_{n,k}^{(j)}}
\over
{\sum_{k=0}^{\infty}ka_{n,k}^{(j)}}}
-(3\mu-1)
{{\sum_{k=0}^{\infty}k^{2}a_{n,k}^{(j)}}
\over
{\sum_{k=0}^{\infty}ka_{n,k}^{(j)}}}
\; \; \; \; .
\eqno (B.10)
$$
Since, $\forall n \geq 2$, there are two values of
$\mu >1$, a further label is necessary to characterize
completely the coupled gauge modes as follows:
$$
g_{1,n}^{(j)}\Bigr(n,\lambda_{1,n}^{(j)},\tau \Bigr)
\; {\rm and} \;
R_{1,n}^{(j)}\Bigr(n,\lambda_{1,n}^{(j)},\tau \Bigr)
\; {\rm if} \;
\mu=\mu_{+}^{(1)}
$$
$$
g_{2,n}^{(j)}\Bigr(n,\lambda_{2,n}^{(j)},\tau \Bigr)
\; {\rm and} \;
R_{2,n}^{(j)}\Bigr(n,\lambda_{2,n}^{(j)},\tau \Bigr)
\; {\rm if} \;
\mu=\mu_{+}^{(2)}
$$
(see (3.20)-(3.21)).

Note that it is extremely difficult (if not impossible)
to find the eigenvalues $\lambda_{n}^{(j)}$ by analytic
or numerical methods, since (B.4)-(B.5) imply that
a function $H$ exists such that
$$
{a_{n,k}^{(j)}\over a_{n,0}^{(j)}}=H(k,n,\mu)
{\Bigr(\lambda_{n}^{(j)}\Bigr)}^{k\over 2}
\eqno (B.11)
$$
for all even values of $k \geq 2$, and
$\forall n \geq 2$. Thus, when (B.11) is inserted
into (B.9)-(B.10), it is not clear how to find an
explicit solution for $\lambda_{n}^{(j)}$ and
$a_{n,k}^{(j)}$.
\vskip 1cm
\leftline {\bf References}
\vskip 1cm
\item {[1]}
Luckock H C and Moss I G 1989
{\it Class. Quantum Grav.} {\bf 6} 1993
\item {[2]}
Poletti S  1990 {\it Phys. Lett.} {\bf 249B} 249
\item {[3]}
D'Eath P D and Esposito G V M 1991 {\it Phys. Rev.} D {\bf 43} 3234
\item {[4]}
D'Eath P D and Esposito G V M 1991 {\it Phys. Rev.} D {\bf 44} 1713
\item {[5]}
Barvinsky A O, Kamenshchik A Y, Karmazin I P and Mishakov I V
1992 {\it Class. Quantum Grav.} {\bf 9} L27
\item {[6]}
Kamenshchik A Y and Mishakov I V 1992 {\it Int. J. Mod. Phys.} A
{\bf 7} 3713
\item {[7]}
Barvinsky A O, Kamenshchik A Y and Karmazin I P 1992
{\it Ann. Phys., N.Y.} {\bf 219} 201
\item {[8]}
Kamenshchik A Y and Mishakov I V 1993 {\it Phys. Rev.}
D {\bf 47} 1380
\item {[9]}
Esposito G 1992 {\it Quantum Gravity, Quantum Cosmology and Lorentzian
Geometries} Lecture
Notes in Physics, New Series m: Monographs vol m12
(Berlin: Springer)
\item {[10]}
Duff M J 1982 Ultraviolet divergences in
extended supergravity {\it Supergravity 1981}
eds S Ferrara and J G Taylor
(Cambridge: Cambridge University Press)
\item {[11]}
D'Eath P D 1986 {\it Nucl. Phys.} B {\bf 269} 665
\item {[12]}
Hawking S W 1977 {\it Commun. Math. Phys.} {\bf 55} 133
\item {[13]}
Branson T P and Gilkey P B 1990 {\it Commun. Part. Diff. Eqns.}
{\bf 15} 245
\item {[14]}
Louko J 1988 {\it Phys. Rev.} D {\bf 38} 478
\item {[15]}
Itzykson C and Zuber J B 1985 {\it Quantum Field Theory}
(New York: McGraw-Hill)
\item {[16]}
Gradshteyn I S and Ryzhik I M 1965 {\it Table of Integrals, Series
and Products} (New York: Academic)
\item {[17]}
Moss I G and Poletti S 1990 {\it Nucl. Phys.} B {\bf 341} 155
\item {[18]}
Moss I G 1989 {\it Class. Quantum Grav.} {\bf 6} 759
\item {[19]}
Greiner P 1971 {\it Archs. Ration. Mech. Analysis} {\bf 41} 163
\item {[20]}
Kennedy G 1978 {\it J. Phys. A: Math. Gen.} {\bf 11} L173
\item {[21]}
Louko J 1991 {\it Class. Quantum Grav.} {\bf 8} L37
\item {[22]}
Louko J 1991 {\it Class. Quantum Grav.} {\bf 8} 1947
\item {[23]}
Louko J 1988 {\it Ann. Phys., N.Y.} {\bf 181} 318
\item {[24]}
Laenen E and van Nieuwenhuizen P 1991
{\it Ann. Phys., N.Y.} {\bf 207} 77
\item {[25]}
Gitman D M and Tyutin I V 1990 {\it Quantization of Fields
with Constraints} (Berlin: Springer)
\item {[26]}
Govaerts J and Troost W 1991
{\it Class. Quantum Grav.} {\bf 8} 1723
\item {[27]}
Govaerts J 1991 {\it Hamiltonian Quantization and Constrained
Dynamics}, Leuven Notes in Mathematical and Theoretical
Physics, Series B, Volume 4 (Leuven: Leuven University Press)
\item {[28]}
Vassilevich D V 1991 {\it Nuovo Cimento} A {\bf 104} 743
\item {[29]}
Vassilevich D V 1992 {\it Nuovo Cimento} A {\bf 105} 649
\item {[30]}
Guven J and Ryan Jr. M P 1992 {\it Phys. Rev.} D {\bf 45} 3559
\item {[31]}
Kunstatter G 1992 {\it Class. Quantum Grav.} {\bf 9} 1469
\item {[32]}
Henneaux M and Teitelboim C 1992 {\it Quantization of Gauge
Systems} (Princeton: Princeton University Press)
\item {[33]}
Vassilevich D V 1993 {\it Int. J. Mod. Phys.} A {\bf 8} 1637
\item {[34]}
Barvinsky A O 1993 {\it Phys. Rep.} {\bf 230} 237
\item {[35]}
Kamenshchik A Y and Mishakov I V 1994
{\it Phys. Rev.} D {\bf 49} 816
\bye